\documentclass[10pt]{article}
\font\smallit=cmti10
\font\smalltt=cmtt10

\usepackage{amssymb,latexsym,amsmath,epsfig,amsthm,array}

 \makeatletter

\renewcommand\section{\@startsection {section}{1}{\z@}
{-30pt \@plus -1ex \@minus -.2ex}
{2.3ex \@plus.2ex}
{\normalfont\normalsize\bfseries}}

\renewcommand\subsection{\@startsection{subsection}{2}{\z@}
{-3.25ex\@plus -1ex \@minus -.2ex}
{1.5ex \@plus .2ex}
{\normalfont\normalsize\bfseries}}

\renewcommand{\@seccntformat}[1]{\csname the#1\endcsname. }

\makeatother

\theoremstyle{definition}
\newtheorem{theorem}{Theorem}[section]
\theoremstyle{plain}

\newtheorem{problem}[theorem]{Problem}

\theoremstyle{definition}
\newtheorem{definition}[theorem]{Definition}

\begin{document}

\begin{center}
\uppercase{\bf The straight line complexity of small factorials and primorials}
\vskip 20pt
{\bf Klas Markstr\"{o}m }\\
{\smallit Department of Mathematics and Mathematical Statistics, Ume\aa{} University,  SE-901 87  Ume\aa, Sweden}\\
{\tt Klas.Markstrom@math.umu.se}\\ 
\vskip 10pt
\end{center}
\vskip 30pt

\centerline{\smallit Received: , Revised: , Accepted: , Published: } 
\vskip 30pt

\centerline{\bf Abstract}

\noindent
	In this paper we determine the straight-line complexity of $n!$
	for $n\leq 22$ and give bounds for the complexities up to $n=46$. 
	In the same way we determine the straight-line complexity of the
	product of the first primes up to $p=23$ and give bounds for
	$p\leq 43$.
	
	Our results are based on an exhaustive computer search of the short length
	straight-line programs.

\pagestyle{myheadings}
\markright{\smalltt INTEGERS: 14 (2014)\hfill}
\thispagestyle{empty}
\baselineskip=12.875pt
\vskip 30pt

\section{Introduction}
In \cite{SS} Shub and Smale studied the complexity of a number of different algebraic problems in terms of the number of 
ring operations needed to compute a given ring element by a \emph{straight line program}.  A straight line program for an integer $y$
can be described as a sequence of tuples $x_k=(x_i\circ x_j)$, where $i\leq j< k$, $x_1=1$, $\circ$ can be any of $+,-,\times$, and the final element $x_f$ is  equal to $y$. The smallest integer $f$ such that there exists a straight line program of length $f$ is called the straight line complexity, or cost, of $y$ and is denoted by $\tau(y)$.

In \cite{BSS} a general complexity theory for computation over rings was introduced, see also \cite{BCSS}, and here the \emph{ultimate complexity} of $n!$ turned out to be of great interest.  For an integer $x$ the ultimate complexity $\tau'(x)$ is defined as the minimum $\tau(y)$ for all $y$ which are integer multiples of $x$.   In particular, if there exists a constant $c$ such that $\tau'(n!)$  is less than $(\log n)^c$ then this would lead to a fast algorithm for factoring integers, see the discussion in \cite{C1,C2}. The non-existence of such a constant $c$ would imply that $P\neq NP$ over the complex numbers \cite{SS} and provide strong lower bounds for several important problems in complexity theory, see \cite{B09} and \cite{Ko}.

The results of \cite{SS,MS,Mo1} provide upper and lower bounds for the straight line complexity of general integers and imply that for most integers $\tau(n)$ is not $O(p(\log\log n))$ for any polynomial $p$. In \cite{LSZ} similar bounds were derived for functions over finite fields. The known bounds for a general integer $n$ are
\begin{equation}\label{bound}
	\log_2(\log_2 n)+1 \leq \tau(n)  \leq 2\log_2 n
\end{equation}
The lower bound is optimal since $\tau(2^{2^k})=k+1$. The upper bound is achieved by first computing the necessary powers of 2 and then adding them according to the binary expansion of $n$.

For specific integers, such as $n!$, there are few results which strengthen the general bounds. However for $n!$ Cheng   derived an improved algorithm, conditional on a conjecture regarding the distribution of smooth integers, and earlier \cite{Str} a weaker, unconditional, bound was derived by Strassen.

The purpose of this short note is to report the \emph{exact} values of $\tau'(n!)$ for small values of $n$ and likewise for $\tau'(p\#)$, where $p\#$ is the primorial, which is the product of all primes less than or equal to $p$. It is easy to see that given a short straight line program for $p\#$ we can also find one for $n!$ by using repeated squaring.
Our results were obtained by first doing an exhaustive computer search of all straight line program up to a given length, followed by an extended search adapted to finding program for $n!$ and $p\#$.

Most of the material in this note was originally part  of a longer paper but while preparing that paper the author found out that the non-computational results were already covered by other recently published papers. That was over ten years ago but given the slow progress on problems in this area we hope that these exact results and bounds will help draw attention to the problems and stimulate interest among new researchers. Additions to the material from the older paper is a recomputation of all data using a newly written program and as a result of this an improvement of some of the lower bounds, and the addition of data for the straight line complexity of the exact factorial and primorials, rather than only multiples of them.

\section{Searching for optimal straight line programs}

\begin{figure}[!ht]
\begin{tabular}{|l|l|l|l|l|}
	\hline
	$k$ & Size of reached set & Initial interval &  Covered interval&Covered set  \\
	\hline
	1          & 2               & 2                & 2     & 2              \\
	2          & 4               & 4                & 4     & 4              \\
         3          & 9               & 6                & 6     & 8              \\	
         4          & 26             & 12             & 12    & 27            \\	
	5          & 102          & 40              & 43    & 125          \\	
         6          & 562          & 112            & 138   & 970          \\		
	7          & 4363        & 310           & 705   & 13384       \\		
         8          & 46154      & 1820         & 3546  & 337096     \\
	9          & 652227   & 10266       & 26686 & 19040788  \\	
	\hline
\end{tabular}
\caption{Statistics for straight line programs of length at most 9}\label{tab1}
\end{figure}
Our bounds have been found by doing an exhaustive search of the set of all straight line programs of a given length.  In Appendix A we give a more detailed description of how the search was performed.   In Figure \ref{tab1} we display some statistics for the straight line programs. We say that an integer $y$ has been reached if there is a straight line program of length at most $k$ which computes $y$, and that $y$ has been covered if $y$ is a divisor of $x_j$ for some $j\leq k$. We also include the length of the longest interval of the form $[1,\ldots,x]$ in which all integers have been reached and covered respectively.

Full data from the search were saved up to $k=9$, after that the space requirements for the full set of programs became prohibitive. For larger $k$ we instead extended the search to higher values of $k$ for specific target integers, in particular the different factorials, primorials and multiples of them. A complete search of this type was made up to $k=11$, thereby finding the optimal program for the cases where the length is at most 11 and providing a lower bound of 12 for the remaining target integers. For certain target integers the complete search could be extended further thanks to the efficiency in pruning the search tree for larger targets, as described in Appendix A.
We also performed searches to extend some heuristically chosen straight line programs, hoping to find  improved upper bounds for some cases.

In Figure \ref{tab2} we show the exact values for $\tau'(n!)$ for $n\leq 28$ and for each such $n$ an example of an optimal straight line program. For larger $n$ we display the best method found by our partial search. The final columns states whether the method is optimal or not, and otherwise the lowest possible value.
In Figure \ref{tab2b} we show the exact values for $\tau(n!)$ for $n \leq 14$, and upper and lower bounds for some larger values of $n$.

Similarly Figure \ref{tab3} and \ref{tab3b} give exact values and bounds for the small primorials, and multiples of them.

The optimal methods are noticeably better than the upper bound for $\tau(n!)$ given in inequality (\ref{bound}). The method of Strassen \cite{Str} gives a bound $\tau(n!)=\mathcal{O}(\sqrt n\log^2{n})$, which seems to deviate more and more from the optimal methods for larger $n$.  The conditional method of Cheng \cite{C1} has a complexity of the form $\mathcal{O}(exp(c\sqrt{\log{n}\log{\log{n}}}))$, which certainly seems compatible with the results for small $n$, but is so sensitive to the value of the constant $c$ that very little can be said based on small values of $n$.

The function $\tau'(n!)$ is a monotone increasing function however it is not obvious that $\tau(n!)$ is. We end this note with an open problem.
\begin{problem}
	Is $\tau(n!)$ a monotone function?
\end{problem}
For small $n$ Table \ref{tab2} shows that $\tau(n!)$ is monotone, but we would not find it surprising if this fails for larger $n$.

\begin{figure}[!hbt]
\begin{tabular}{|p{0.3cm}|p{0.3cm}|p{7cm}|p{1cm}|}
	\hline
	$n$ & f & Program &  Lower bound\\
	\hline
	2  & 1  & $ \{1,1,+\}  $& Opt \\
	\hline
	3  & 3  & $\{1, 1,+\}, \{1, 2,+\}, \{2, 3,*\}$& Opt \\
	\hline
	4  & 4  & $\{1, 1,+\}, \{2, 2,+\}, \{2, 3,+\},\{3, 4,*\}$& Opt \\
	\hline
	5  & 5  & $\{1, 1,+\}, \{2, 2,+\}, \{3,3,*\},\{4,1,-\},$ $\{4, 5,*\}$& Opt \\
	\hline
	6-7  & 6  & $\{1, 1,+\}, \{2, 2,*\}, \{3,3,*\},\{4,4,*\},$ $\{5, 5,*\},\{6, 4,-\}$& Opt \\
	\hline
	8-10  & 7   & $\{1, 1,+\}, \{2,2,+\}, 	\{3,3,*\},\{4,4,*\},$ $\{5, 5,*\},\{6, 4,-\},\{7, 7,*\}$& Opt \\
	\hline
	11-14  & 9  & $\{1, 1,+\}, \{2,2,+\}, 	\{3,3,*\},\{4,4,*\},$ $\{5, 3,+\},\{6, 4,*\},\{7, 2,-\},\{7,8,*\},$	$\{9, 9,*\}$& Opt \\
	\hline
	15-17  & 10 & $\{1, 1,+\}, \{2,2,+\},\{3,3,*\},\{4,4,*\},${\ }$\{5, 5,*\}, 	\{6,6,*\},\{5,7,-\},\{8,8,*\},${\ }$\{8,9,-\},\{9, 10,*\}$&  Opt\\
	\hline
	18-19  & 11   & $\{1, 1,+\}, \{2,2,+\},\{3,3,*\},\{4,2,+\},${\ }$\{5, 5,*\},\{6, 4,-\}, \{6,7,*\},\{6,8,*\},${\ }$\{9,7,-\},\{9, 10,*\},\{11, 11,*\}$&  Opt\\
	\hline
	20-22  & 12  & $\{1, 1,+\},\{2,2,+\},\{3,3,*\},\{4,4,*\},${\ }$\{3, 5,+\},\{6, 4,*\}, 	\{2,7,-\},\{7,8,*\},${\ }$\{9,9,*\},\{10, 5,-\},\{10, 11,*\},${\ }$\{10, 12,*\}$& Opt \\
	\hline
	23-28  & 14  & $\{1, 1,+\},
	\{2,2,*\},\{3,3,*\},\{4,4,*\},${\ }$\{5, 5,*\},\{6, 6,*\}, 	\{5,7,*\},\{8,4,-\},${\ }$\{8,9,*\},\{10, 9,-\},\{8, 11,+\},${\ }$\{10,12,*\},\{13,13,*\},\{14,14,*\}$, & Opt \\
	\hline
	29-34  & 16   & $\{1, 1,+\},
	\{2,2,*\},\{3,3,*\},\{4,4,*\},${\ }$\{5, 5,*\},\{6, 6,*\}, 	\{5,7,*\},\{8,4,-\},${\ }$\{8,9,*\},\{10, 9,-\},\{8, 11,+\},\{10,12,*\}, ${\ }$\{13,13,*\},\{14,14,*\}$,\{7,4,-\}$,\{14,15,*\}$ & 14 \\
	\hline
	35-46  & 17  & 
	$\{1, 1,+\},\{2,2,*\},\{3,3,*\},\{4,4,*\},\{5, 5,*\}$ {\ }  $\{6, 6,*\},	\{5,7,*\},\{8,4,-\},\{8,9,*\}${\ }$\{10, 9,-\},\{10, 11,+\},
	\{11,12,*\},\{13,6,-\}$ {\ }$\{11,14,*\},\{15,15,*\},\{16,16,*\},\{17,17,*\}$ 
	& 14  \\		
	\hline
\end{tabular}
\caption{Straight line programs for multiples of $n!$}\label{tab2}
\end{figure}

\begin{figure}[!hbt]
\begin{tabular}{|p{0.3cm}|p{0.3cm}|p{7cm}|p{1cm}|}
	\hline
	$n$ & f & Program &  Lower bound\\
	\hline
	2  & 1  & $ \{1,1,+\}  $& Opt \\
	\hline
	3  & 3  & $\{1, 1,+\}, \{1, 2,+\}, \{2, 3,*\}$& Opt \\
	\hline
	4  & 4  & $\{1, 1,+\}, \{2, 2,+\}, \{2, 3,+\},\{3, 4,*\}$& Opt \\
	\hline
	5  & 6  & $\{1, 1,+\}, \{1, 2,+\}, \{1,3,+\},\{3,4,*\},$ $\{5,2,-\},\{5,6,*\}$& Opt \\
	\hline
	6  & 6  & $\{1, 1,+\}, \{1, 2,+\}, \{3,3,*\},\{3,4,*\},$ $\{5, 5,*\},\{6, 4,-\}$& Opt \\
	\hline
	7  & 7  & $\{1, 1,+\}, \{1, 2,+\}, \{2,3,*\},\{2,4,*\},$ $\{4, 5,*\},\{6, 2,-\},\{6, 7,*\}$& Opt \\
	\hline
	8  & 8   & $\{1, 1,+\}, \{1,2,+\}, \{1,3,+\},\{3,4,*\},$ $\{5, 5,*\},\{6, 4,-\},\{6, 7,*\},\{8, 2,-\}$& Opt \\	
	\hline
	9  & 8   & $\{1, 1,+\}, \{1, 2,+\}, \{2,3,*\},\{2,4,*\},$ $\{4, 5,*\},\{6, 2,-\},\{6, 7,*\},\{7, 8,*\}$& Opt \\	
	\hline
	10  & 9   & $\{1, 1,+\}, \{1, 2,+\}, \{2,3,+\},\{2,4,+\},$ $\{4, 5,+\},\{6, 6,*\},\{4, 7,*\},\{5, 8,*\},\{8, 9,*\}$& Opt \\	
	\hline
	11  & 9   & $\{1, 1,+\}, \{2, 2,+\}, \{3,3,*\},\{3,4,+\},$ $\{4, 5,*\},\{3, 6,+\},\{5, 7,*\},\{8, 6,-\},\{8, 9,*\}$& Opt \\	
	\hline
	12  & 10   & $\{1, 1,+\}, \{2, 2,+\}, \{3,3,*\},\{2,4,+\}, \{4, 5,*\},$ $\{4, 6,+\},\{7, 7,*\},\{8, 4,-\},$ $\{6, 9,*\},\{5, 10,*\}$& Opt \\	
	\hline
	
	13 & 11   & $\{1, 1,+\}, \{1, 2,+\}, \{1,3,+\},\{3,3,*\},\{4, 5,*\},$ $\{3, 6,+\},\{5, 6,*\},\{7, 8,*\},\{7, 9,*\},\{10, 4,-\},$ $\{9, 11,*\}$ & Opt \\		
	\hline
	14 & 11 & $\{1, 1,+\}, \{2, 2,*\}, \{3,3,*\},\{3,4,+\},\{2, 5,+\},$ $\{5, 6,+\},\{5, 7,*\},\{6, 8,*\},\{4, 9,*\},\{10, 8,-\},$ $\{10, 11,*\}$   & Opt \\		
	\hline
	
	15 &  12  & $\{1, 1,+\}, \{2, 2,*\}, \{2,3,*\},\{1,4,+\},\{3, 5,*\},$ $\{6, 4,-\},\{6, 7,*\},\{8, 4,-\},\{6, 9,*\},\{10, 6,+\},$ $\{8, 11,*\},\{10, 12,*\}$   & Opt \\		
	\hline
	16 &  12  & $\{1, 1,+\}, \{2, 2,*\}, \{2,3,+\},\{3,4,+\},\{3, 4,*\},$ $\{4, 5,*\},\{6, 7,+\},\{6, 7,*\},\{9, 5,-\},\{8, 9,*\},$ $\{10, 11,*\},\{11, 12,*\}$  & Opt \\		
	\hline
	17 &  12  & $\{1, 1,+\}, \{2, 2,*\}, \{2,3,+\},\{2,4,*\},\{5, 5,*\},$ $\{6, 3,-\},\{5, 6,+\},\{6, 7,*\},\{8, 9,*\},\{4, 10,*\},$ $\{11, 9,-\},\{11, 12,*\}$  & Opt \\		
	\hline

	18 &  13  & $\{1, 1,+\}, \{1, 2,+\}, \{2,3,*\},\{3,4,*\},\{3, 5,+\},$ $\{6, 6,*\},\{5, 7,+\},\{6, 8,+\},\{5, 9,*\},\{7, 10,*\},$ $\{7, 11,*\},\{12, 10,-\},\{11, 13,*\}$  & Opt \\		
	\hline
	19 &  13  & $\{1, 1,+\}, \{2, 2,+\}, \{3,3,*\},\{4,2,-\},\{4, 2,+\},$ $\{4, 5,*\},\{7, 3,-\},\{6, 6,*\},\{7, 9,*\},\{9, 10,*\},$ $\{11, 7,-\},\{11, 12,*\},\{8, 13,*\}$  & Opt \\		
	\hline
	20 &  14  & $\{1, 1,+\}, \{2, 2,+\}, \{3,3,*\},\{1,4,+\},\{3, 5,*\},$ $\{6, 4,-\},\{7, 7,*\},\{8, 3,-\},\{8, 4,-\},\{5, 9,+\},$ $\{5, 11,*\},\{9, 10,*\},\{13, 13,*\},\{12, 14,*\}$  & 13 \\		
	\hline

	\hline
\end{tabular}
\caption{Straight line programs for  $n!$}\label{tab2b}
\end{figure}

\begin{figure}[!ht]
\begin{tabular}{|p{0.3cm}|p{0.3cm}|p{6cm}|p{1cm}|}
	\hline
	$p$ & f  & Program &  lower bound\\
	\hline
	2  & 1  & $ \{1,1,+\}  $& Opt \\
	\hline
	3  & 3 & $\{1, 1,+\}, \{1, 2,+\}, \{2, 3,*\}$& Opt \\
	\hline
	5  & 5 & $\{1, 1,+\}, \{2, 2,*\}, \{3,3,*\},\{4,4,*\},$ {\ } $\{4,5,-\}$& Opt \\
	\hline	
	7  & 6  & $\{1, 1,+\}, \{2, 2,*\},	\{3,3,*\},\{4,4,*\},$ {\ } $\{5,5,*\},\{4,6,-\}$& Opt \\
	\hline		
	11  & 7  & $\{1, 1,+\}, \{2, 2,*\},	\{3,3,*\},\{3,4,*\},$ {\ } $\{3,5,+\},\{6,6,*\},\{3,7,-\}$& Opt \\
	\hline		
	13  & 8  & $\{1, 1,+\}, \{2, 2,*\},	\{3,3,*\},\{4,4,*\},$ {\ } $\{5,5,*\},\{6,6,*\},\{7,7,*\},\{4,8,-\}$& Opt \\
	\hline
	17  & 9  & $\{1, 1,+\}, \{2,2,*\},\{3,3,*\},\{4,4,*\},$ {\ } $\{5,5,*\},\{6,6,*\},\{7,7,*\},\{8,8,*\},$ {\ } $\{9,5,-\}$& Opt \\
	\hline
	19-23  & 10 & $\{1, 1,+\},  \{2,2,*\},\{3,3,*\},\{4,4,*\},$ {\ }$\{5,2,-\},\{6,6,*\},\{7,7,*\},\{8,8,*\}, 	$ {\ } $\{9,9,*\},\{10,8,-\}$ & Opt \\
	\hline
	29-31  & 11 & $\{1, 1,+\},  \{1,2,+\},\{2,3,+\},\{2,4,*\},$ {\ }$\{5,5,*\},\{6,6,*\},\{7,4,+\},\{7,5,+\}, 	$ {\ } $\{9,3,+\},\{9,8,*\},\{11,10,*\}$ & Opt \\
	\hline
	37-43  & 14  & $\{1, 1,+\},\{2,2,*\},\{3,3,*\},\{4,4,*\},$ {\ }  $\{5, 5,*\}\{6, 6,*\}, \{5,7,*\},\{8,4,-\},$ {\ }  $\{8,9,*\},\{10, 9,-\},\{10, 11,+\},$ {\ }  $
	\{11,12,*\},\{13,6,-\},\{11,14,*\}$ & 13 \\			
	\hline
	
\end{tabular}
\caption{Straight line programs for multiples of $p\#$}\label{tab3}
\end{figure}

\begin{figure}[!ht]
\begin{tabular}{|p{0.3cm}|p{0.3cm}|p{6cm}|p{1cm}|}
	\hline
	$p$ & f  & Program &  lower bound\\
	\hline
	2  & 1  & $ \{1,1,+\}  $& Opt \\
	\hline
	3  & 3 & $\{1, 1,+\}, \{1, 2,+\}, \{2, 3,*\}$& Opt \\
	\hline
	5  & 5 & $\{1, 1,+\}, \{1, 2,+\}, \{2,3,+\},\{2,3,*\},$ {\ } $\{4,5,*\}$& Opt \\
	\hline	
	7  & 6  & $\{1, 1,+\}, \{1, 2,+\}, \{2,3,+\},\{3,4,*\},$ {\ } $\{5,1,-\},\{5,6,*\}$& Opt \\
	\hline		
	11  & 7  & $\{1, 1,+\}, \{1, 2,+\},	\{2,3,*\},\{2,4,+\},$ {\ } $\{4,5,*\},\{6,6,*\},\{4,7,+\}$& Opt \\
	\hline		
	13  & 8  & $\{1, 1,+\}, \{1, 2,+\},	\{2,3,+\},\{2,4,*\},$ {\ } $\{5,5,*\},\{6,6,*\},\{5,7,+\},\{3,8,*\}$& Opt \\
	\hline
	17  & 9  & $\{1, 1,+\}, \{2,2,*\},\{3,3,*\},\{1,4,+\},$ {\ } $\{3,5,+\},\{2,5,*\},\{6,7,*\},\{1,8,+\},$ {\ } $\{8,9,*\}$& Opt \\
	\hline
	19 & 10 & $\{1, 1,+\},  \{1,2,+\},\{2,3,*\},\{4,4,*\},$ {\ }$\{4,5,*\},\{6,4,-\},\{6,1,-\},\{8,8,*\}, 	$ {\ } $\{9,5,-\},\{10,7,*\}$ & Opt \\		
	\hline
	23 & 11  &$\{1, 1,+\},  \{1,2,+\},\{2,3,+\},\{2,4,*\},$ {\ }$\{5,3,+\},\{5,6,*\},\{5,7,*\},\{5,8,+\}, $ {\ } $\{9,9,*\},\{10,1,-1\},\{11,7,*\}$      & Opt \\		
	\hline
	29  & 13   &  $\{1, 1,+\},  \{2,2,+\},\{3,3,*\},\{2,4,+\},$ {\ }$\{1,5,+\},\{2,6,*\},\{7,7,*\},\{6,8,+\}, $ {\ } $\{4,9,+\},\{4,10,+\},\{2,9,*\},\{10,11,*\},${\ }$\{12,13,*\}$      & 12 \\			
	\hline
	31 & 15  & $\{1, 1,+\},  \{2,2,+\},\{3,3,*\},\{2,4,+\},$ {\ }$\{1,5,+\},\{2,6,*\},\{7,7,*\},\{6,8,+\}, $ {\ } $\{4,9,+\},\{4,10,+\},\{2,9,*\},\{10,11,*\},${\ }$\{12,13,*\},\{4,1,-\},\{14,15,*\}$    & 12 \\			

	\hline

\end{tabular}
\caption{Straight line programs for $p\#$}\label{tab3b}
\end{figure}

\section*{Acknowledgements}
This research was conducted using the resources of High Performance Computing Center North (HPC2N).  The author would like to thank Charles R Greathouse and Rich Schroeppel for pointing out an error in the first version of the paper, and the anonymous referee for constructive criticism.


\newpage

\section*{Appendix A}\label{app}
Our bounds have been found by doing a two stage search of the set of all straight line programs of a given length. 

\begin{definition}
	A straight line program is \emph{normalized} if
	\begin{enumerate}
		\item $x_i \neq x_j$ if $i\neq j$		
		\item $x_i>0$ for all $i$
 	\end{enumerate}
\end{definition}
It is easy to see that an optimal straight line program for an integer $n$ must satisfy 1, and that every $n$ has an optimal straight line program which satisfies 2.

Further we say that two straight line programs $p_1$ and $p_2$, both of length $k$, are \emph{range-isomorphic} if the sequence of number computed by $p_2$ is a permutation of the sequence computed by $p_1$.  It is easy to see that this is an equivalence relation on the set of straight line programs.

Our search for optimal straight line programs was performed in two stages. First we found one representative for each range-isomorphism equivalence class of the   normalized straight line programs of length up to $k=9$. Second, a search targeted at specific integers was performed.

The first stage was done as follows, starting from the initial straight line program just containing the number 1. 
\begin{enumerate}
	\item Increase $k$ by 1 and continue. 
	\item Extend all programs of length $k-1$ by one step in every possible way. Discard those of the resulting programs which are not normalized.
	\item Reduce the set of all programs of length $k$ by only keeping one representative for each range-isomorphism equivalence class.  
	\item Repeat from 1.
\end{enumerate}
Step 3 is  done since if one replaces an initial segment $p_0$, of length $t$, of a straight line program $p_1$ by a range-isomorphic straight line program $p_0'$ then we can modify, by changing some of the indices, the resulting program to a new program $p_1'$ which computes the same set of numbers as $p_1$.  So, if $p_1$ was an optimal program for some integer $N$ then $p_1'$ is also optimal for $N$.

After stage 1 is done we have found optimal programs for all integers $N$ with $\tau(N)\leq 9$, and have shown that $\tau(N)\geq 10$ for all other integers.

After the set of programs of length 9 had been found in this way we went on with the second stage search. For each target integer  $N$, such that $\tau(N)\geq 10$ each program of length 9 was extended in a depth first search. 

Given a  target integer $N$,  each straight line program of length 9, found in the first stage search, was recursively extended  by one operation, up to a specified maximum length $K$, with the following pruning criteria.
\begin{enumerate}
	\item If the current program $p$  computes the target $N$ then save the program and do not extend it further. 
	\item If the current program $p$ is not normalized then do not extend it further.
	\item If the current program has length $k$ and the maximum integer $x$ which it has computed satisfies $x^{2^{(K-k)}}< N$ then do not extend it further.
\end{enumerate}
The third condition is included since if a program $p$ of this type is extended by $K-k$ steps then the resulting program cannot compute an integer as large as the target $N$, if $k\geq 2$.

Using this depth first search strategy each program of length 9 was extended to $k=11$ for each of our target integers.  For the larger target integers the search could be completed for larger values of $k$ as well, thanks to the more restrictive bound in the third pruning criterion, thus providing larger lower bounds for the optimal straight line  programs, and proving the optimality of the some of the programs found.

\end{document}